%% file: LargeDE.tex
\documentclass[aps,prb,amsmath,showpacs,amssymb,amsfonts,10pt]{revtex4-1}

\usepackage{graphicx}
\usepackage{subfigure}

\input{com.tex}

\begin{document}

\title{Novel approach to description of quantum magnets with large singe-ion easy-plane anisotropy} 
\author{A.V. Sizanov$^1$}
\email{alexey.sizanov@gmail.com}
\author{A.V. Syromyatnikov$^{1,2}$}
\email{syromyat@thd.pnpi.spb.ru}
\affiliation{$^1$Petersburg Nuclear Physics Institute, Gatchina, St.\ Petersburg 188300, Russia}
\affiliation{$^2$Department of Physics, St.\ Petersburg State University, 198504 St.\ Petersburg, Russia}

\date{\today}

\begin{abstract}

We introduce a new representation of an integer spin $S$ via bosonic operators which is useful in describing the paramagnetic phase and transitions to magnetically ordered phases in magnetic systems with large single-ion easy-plane anisotropy $D$. Considering the exchange interaction between spins as a perturbation and using the diagram technique we derive the elementary excitation spectrum and the ground state energy in the third order of the perturbation theory. In the special case of $S=1$ we obtain these expressions also using simpler spin representations some of which were introduced before. Comparison with results of previous numerical studies of 2D systems with $S=1$ demonstrates that our approach works better than other analytical methods applied before for such systems. We apply our results for the elementary excitation spectrum analysis obtained experimentally in $\rm NiCl_2$-$\rm 4SC(NH_2)_2$ (DTN). It is demonstrated that a set of model parameters (exchange constants and $D$) which has been used for DTN so far describes badly the experimentally obtained spectrum. A new set of parameters is proposed using which we fit the spectrum and values of two critical fields of DTN.

\end{abstract}


\maketitle

\section{Introduction} 
\label{secIntro}

Many analytical approaches to consideration of quantum magnetic systems with localized spins base on representations of spins via bosonic or fermionic operators allowing to turn from spin Hamiltonians to those describing ensembles of bosons or fermions. \cite{auer} At low temperature the form of the representation depends on the ground state of the spin system under discussion. The Holstein-Primakoff and the Dyson-Maleev representations are frequently used for magnetically ordered phases, the Jordan-Wigner transformation proved to be useful for $S=1/2$ spin chains, the bond-operator formalism \cite{18} was proposed for spin liquids with singlet ground state, etc. 

We propose in the present paper a new representation of an integer spin $S$. This representation should be useful in describing the paramagnetic phase in which all spins are mainly in the quantum state with zero quantum number for projection onto a preferential direction. As an example of particular system of this type we choose a Heisenberg magnet with large single-ion easy-plane anisotropy which Hamiltonian has the form
\begin{eqnarray}
\label{ham}
 {\cal H} = \half \sum_{i,j} J_{ij} \mathbf{S}_i \mathbf{S}_j + D \sum_i \left(S^z_i\right)^2,
\end{eqnarray}
where summations are taken over all sites of the lattice with arbitrary spatial dimension, $D>0$ and signs of exchange constants $J_{ij}$ are not important in the following. The first term in \eqref{ham} is supposed to be small enough compared to the second one so that the system to be in the paramagnetic phase at $T=0$. There is a quantum critical point (QCP) $D=D_{c}$ separating the paramagnetic phase at $D>D_{c}$ and a magnetically ordered or a spin-liquid one at $D<D_{c}$. The value of $D_{c}$ as well as the phase type at $D<D_{c}$ depend on the lattice spatial dimension and the exchange interaction \cite{add5, 6, wong}. 

Quite a few compounds can be mentioned with the paramagnetic ground state of the considered type which are described by the Hamiltonian \eqref{ham}: $\rm CsFeBr_3$ \cite{add9,lind2,add18,add19}, $\rm CsFeCl_3$ \cite{lind2}, $\rm Sr_3NiPtO_6$ \cite{add1}, NENC \cite{add7,add16,add17}, NENP \cite{add13,add14,add15}, NBYC \cite{add10}, $\rm NiCl_2$-$\rm 4SC(NH_2)_2$ \cite{12,13,14,15,16,17,add8,chern} (apparently the most intensively studied compound of this type in recent years), $\rm(Ni(C_5H_5NO)_6)(NO_3)_2$ \cite{carlin} and $\rm NiSnCl_6\cdot6H_2O$ \cite{nisn1,nisn2}. All the mentioned materials have $S=1$ and all of them are quasi-1D magnets. It is probably the reason why the majority of the recent theoretical investigations of the model \eqref{ham} with large $D>0$ focus on weakly coupled or independent spin chains \cite{add2,add3,add4,add5,add6,add11,add12,add15,kolezh,mc1,mc2,1,4,5,6,7,8,9}. In particular, the elementary excitation spectrum has been derived before using a random phase approximation \cite{lind1,lind2}, the regular perturbation theory \cite{4}, a "generalized spin-wave approach" \cite{13,chern} and some other self-consistent procedures \cite{1,3,5,wang5}. 

Treating the first term in Eq.~\eqref{ham} as a perturbation, using the proposed spin representation and the conventional diagram technique we derive below expressions for the elementary excitation spectrum and the ground state energy in the paramagnetic phase in the third order in the perturbation theory. For brevity, this approach is referred to hereafter as an expansion in terms of $J/D$ while one should bare in mind that constants $J_{ij}$ in Eq.~\eqref{ham} are assumed to be arbitrary. We also obtain these results in the special case of $S=1$ using simpler spin representations some of which were introduced before in Refs.~\cite{kolezh,3}. In the particular case of a spin chain with $S=1$ our expression for the spectrum coincides with that obtained in Ref.~\cite{4} only for this special case using the regular perturbation theory. Comparing our results with those of numerical calculations \cite{2,3} which were carried out for $S=1$, square lattice and antiferromagnetic exchange we show that our approach works better than other theoretical methods proposed so far \cite{1,lind1,5,chern} . In particular, our results are in very good agreement with the numerical ones \cite{2,3} not very close to the QCP. At $D\approx D_c$ only the spectrum of long-wavelength excitations is reproduced unsatisfactorily that is a consequence of strong fluctuations near the QCP.

We demonstrate that our approach is applicable to the intensively studied compound $\rm NiCl_2$-$\rm 4SC(NH_2)_2$ (DTN) \cite{12,13,14,15,16,17,add8,chern} described by the model \eqref{ham}. It is shown that our expression for the spectrum describes badly the spectrum obtained in the neutron experiment \cite{13} with the conventional set of parameters (values of exchange constants and $D$) attributed to DTN before. A new set of parameters is proposed for DTN using which we fit well the neutron data and reproduce values of critical fields found experimentally \cite{16}. It should be noted that in comparison with the conventional model used for DTN discussion we take into account also an exchange interaction between two magnetic sublattices of DTN \cite{prev} which should be taken into consideration as the recent ESR experiment \cite{12} demonstrates.

The rest of the present paper is organized as follows. In Sec.~\ref{spinrep} the new representation of an integer spin is proposed and details of the diagram technique based on this representation are discussed. In Sec. \ref{observ} expressions for the elementary excitation spectrum and the ground state energy are derived in the third order in $J/D$. In Sec.~\ref{disc} we compare our results with those obtained before using other approaches. We discuss also representations which are valid only for $S=1$ some of which are simpler than the general one. The expression for the spectrum derived in Sec.~\ref{observ} is applied for analysis of the DTN spectrum obtained in the neutron experiment \cite{13}. Sec.~\ref{conc} contains our conclusion.

\section{Representation of an integer spin and technique}
\label{spinrep}

The ground state of a system described by the Hamiltonian \eqref{ham} in the limit of $J/D \to 0$ is a direct product of states $| S^{z}_i = 0 \rangle$: $\Pi_i \otimes | S^z_i = 0 \rangle$. The lowest excited states can be constructed from the ground state by substituting $| S^z_i = \pm 1 \rangle$ for $| S^{z}_i = 0 \rangle$ at any $i$. The energy of such states is equal to $D$ and one leads to the doubly degenerate dispersionless elementary excitation spectrum
\begin{eqnarray}
\label{spec00}
 \e_{0\pp} = D.
\end{eqnarray}
The exchange interaction gives rise to the spectrum dispersion. When it is small enough one can find expressions for the spectrum and other observables considering the exchange as a perturbation. In particular, the elementary excitation spectrum of the spin chain with $S=1$ and the exchange coupling between only nearest neighbors was calculated in Ref.~\cite{4} in the third order of the regular (non diagrammatic) perturbation theory. In contrast, our aim is to construct a spin representation which opens the door to calculations utilizing the diagram technique and which allows to obtain such expressions for observables easier. In particular, we recover below the result of Ref.~\cite{4} using the diagram technique. 

We propose the following expressions for projections of an integer spin $S$:
\begin{eqnarray}
 \label{sz} 
 S^z_i &=& n_{b,i} - n_{a,i}, \\
\label{s+} 
S^+_i &=& S^x_i + i S^y_i = \bkr_i \sqrt{\dfrac{ ( S - n_{b,i} ) ( S+1 + n_{b,i}) }{ 1 + n_{b,i} }} +
		\sqrt{\dfrac{ ( S - n_{a,i} ) ( S+1 + n_{a,i}) }{ 1 + n_{a,i} }} \cdot a_i,  
\end{eqnarray}
where $a_i$ and $b_i$ are bosonic operators, $n_{a,i} = \akr_i a_i$, $n_{b,i} = \bkr_i b_i$ and
\begin{eqnarray}
\langle S^z_i = -(n+1) | \;\, \akr_i \;\, | S^z_i = -n \rangle  &\;=& \sqrt{ n+1 },  \non \\
\langle S^z_i = n+1 | \;\, \bkr_i \;\, | S^z_i = n \rangle &\;=& \sqrt{ n+1 },  \non \\
 a_i \; | S^z_i = 0 \rangle &\;=& 0, \non\\ 
 b_i \; | S^z_i = 0 \rangle &\;=& 0, \non \\
 n &\geq& 0. \non
\end{eqnarray}
Operators $\akr$ and $\bkr$ create excitations with $S^z=-1$ and $+1$, respectively. The subspace of physical states is constrained by the following conditions:
\begin{equation}
\label{consab} 
n_{a,i} n_{b,i} = 0, 
\end{equation}
\begin{equation}
\label{consaa}
\begin{array}{l}
 n_{a,i} \leq S,  \\
 n_{b,i} \leq S. 
\end{array}
\end{equation}
It can be readily verified that representation \eqref{sz}--\eqref{s+} reproduces the spin commutation relations on the physical subspace defined by Eqs.~\eqref{consab} and \eqref{consaa}.

Condition \eqref{consab} selects states having at any site only excitations of $a$- or $b$- type. This constraint can be satisfied by adding to the Hamiltonian the term describing an infinite repulsion between $a$ and $b$ particles
\begin{equation}
\label{u1} 
{\cal H}_{U1} = \frac UN \sum_i  \akr_i \bkr_i a_i b_i, \quad U \to + \infty. 
\end{equation}
After this modification matrix elements of operators \eqref{sz}--\eqref{s+} become zero between states from physical and unphysical subspaces constrained by conditions \eqref{consaa}. It means that at zero (and most probably at low) temperature we can use Eqs.~\eqref{sz}--\eqref{s+} and the diagram technique forgetting about condition \eqref{consaa}, as it is done in a similar situation in the case of the Holstein-Primakoff representation. We prove this statement below for $S=1$ by performing calculations using Eqs.~\eqref{sz}--\eqref{s+}, taking into account constraint term \eqref{u1} and introducing to the Hamiltonian the additional term 
\begin{eqnarray}
\label{u2} 
{\cal H}_{U2} = \frac UN \sum_i \left( \akr_i \akr_i a_i a_i + \bkr_i \bkr_i b_i b_i \right), \quad U \to + \infty,  
\end{eqnarray}
which explicitly selects states with no more than one $a$ or $b$ particles as condition \eqref{consaa} requires at $S=1$. It is worth noting that we could construct a spin representation similar to \eqref{sz}--\eqref{s+}, which matrix elements are zero on states from physical and unphysical subspaces so that it was not necessary to introduce term \eqref{u1} to the Hamiltonian. However such a representation would be very cumbersome. On the other hand term \eqref{u1} does not complicate calculations much at any integer $S$. That is why we use below Eqs.~\eqref{sz}--\eqref{s+} with constraint term \eqref{u1}. 

At sufficiently small exchange interaction and low temperature we expect densities of $a$ and $b$ particles to be small. Therefore one can expand square roots in Eq.~\eqref{s+} into series and restrict oneself by the first terms of the resultant normally ordered expressions which have the form
\begin{eqnarray}
\label{s+series}
 S^+_i &\approx& \bkr_i \left( c_1 - c_2 \; \bkr_i b_i \right) + \left( c_1 - c_2 \; \akr_i a_i \right) a_i, \\
 \label{c1}
 c_1 &=& \sqrt{ S(S+1) }, \\
 \label{c2}
 c_2  &=& \sqrt{ S(S+1) } - \sqrt{ \frac{(S-1)(S+2)}{2} } > 0.
\end{eqnarray}
Using Eqs.~\eqref{sz} and \eqref{s+series} and taking into account Eq.~\eqref{u1} we obtain from Eq.~\eqref{ham} 
\begin{subequations}
\label{hamfur}
\begin{eqnarray}
\label{h2}
{\cal H} &=& \sum_\pp \e_{1\bf p} \left( \akr_\pp a_\pp + \bkr_\pp b_\pp \right) \\
\label{h2an}
&&{}+\sum_\pp \frac{ c_1^2 }{ 2 } J_\pp \left( \akr_\pp \bkr_{-\pp} + a_\pp b_{-\pp} \right) \\
\label{aaaa} 
&&{} + \fracNN \sum_{\pp_1+\pp_2=\pp_3+\pp_4} \left\{ \left[ D + \half J_{3-1} - \frac{ c_1 c_2 }{ 2 } \left( J_1 + J_3 \right) \right] \left( \akr_1 \akr_2 a_3 a_4 + \bkr_1 \bkr_2 b_3 b_4 \right) + \left[ U - J_{3-1} \right] \akr_1 \bkr_2 a_3 b_4  \right\} \\
\label{abbb}
&&{} - \fracNN \sum_{\pp_1+\pp_2+\pp_3=\pp_4} \frac{ c_1 c_2 }{ 2 } J_1 \left( \bkr_1 \akr_2 \akr_3 a_4 + \akr_1 \bkr_2 \bkr_3 b_4 + \akr_4 a_3 a_2 b_1 + \bkr_4 b_3 b_2 a_1 \right),
\end{eqnarray}
\end{subequations}
where $ J_\pp = \sum_j J_{ij} e^{i \pp {\bf R}_{ij}} $, $N$ is the number of unit cells,
\beq
\label{spec0}
\e_{1\bf p} = \e_{0\bf p} + \frac{ c_1^2 }{ 2 } J_\pp = D + \frac{ c_1^2 }{ 2 } J_\pp,
\eeq
$\e_{0\bf p}$ is defined by Eq.~\eqref{spec00} and we omit indexes $\pp$ in Eqs.~\eqref{aaaa} and \eqref{abbb}. Note that we take into account only terms with no more than four operators in Eq.~\eqref{hamfur}. It can be shown that terms containing more than four operators which appear from higher order terms in the series expansion of square roots in Eq.~\eqref{s+} lead to contributions to the spectrum and to the ground state energy of the order of $(J/D)^4$ and higher. As our aim is to calculate these quantities only up to the third order in $J/D$, we can use Hamiltonian \eqref{hamfur}.

It is convenient to introduce the following Green's functions:
\begin{eqnarray}
\label{gdef}
 G ( p ) &=& - i \langle a_p \akr_p \rangle = - i \langle b_p \bkr_p \rangle, \\
 \label{fdef}
 F ( p ) &=& - i \langle \bkr_{-p} \akr_p \rangle,
\end{eqnarray}
where $ p = (\w , \pp ) $ and $ a_p $ is the Fourier transform of $ a_\pp ( t ) $. Naturally, the equality $\langle a_p \akr_p \rangle =  \langle b_p \bkr_p \rangle$ is satisfied. Dyson equations for these Green's functions have the form
\begin{eqnarray}
\label{dyson}
 G ( p ) &=& G_0 ( p ) \left[ 1 + \Sig_p G ( p ) + \Pi_p F ( p ) \right], \\
 F ( p ) &=& G_0 ( -p) \left[ \overline{\Pi}_p G ( p ) + \Sig_{-p} F ( p ) \right],\non
\end{eqnarray}
where $ G_0 ( p ) = ( \w - \e_{1\bf p}+ i\delta )^{-1} $, $\Sig_p$ and $\Pi_p$ are normal and anomalous self-energy parts, respectively. The solution of Eq.~\eqref{dyson} has the form
\begin{eqnarray}
 \label{g}
 G ( p ) &=& \frac{\w+ \e_{1\bf p} + \Sig_{-p}}{{\cal D}( p )}, \\
 \label{f}
 F ( p ) &=& - \frac{ \overline{\Pi}_p}{{\cal D}( p )}, \\
 \label{den} 
 {\cal D}( p ) &=& \w^2 - \e^2_{1\bf p} - \e_{1\bf p} (\Sig_p + \Sig_{-p}) + \w ( \Sig_{-p} - \Sig_p ) - \Sig_p \Sig_{-p} + \left|\Pi_p\right|^2.
\end{eqnarray}

\section{Application of the approach}
\label{observ}

In this section we apply the method described above for calculation of the elementary excitation spectrum and the ground state energy.

\subsection{Elementary excitation spectrum}

\begin{figure}
\includegraphics[scale = 1.0]{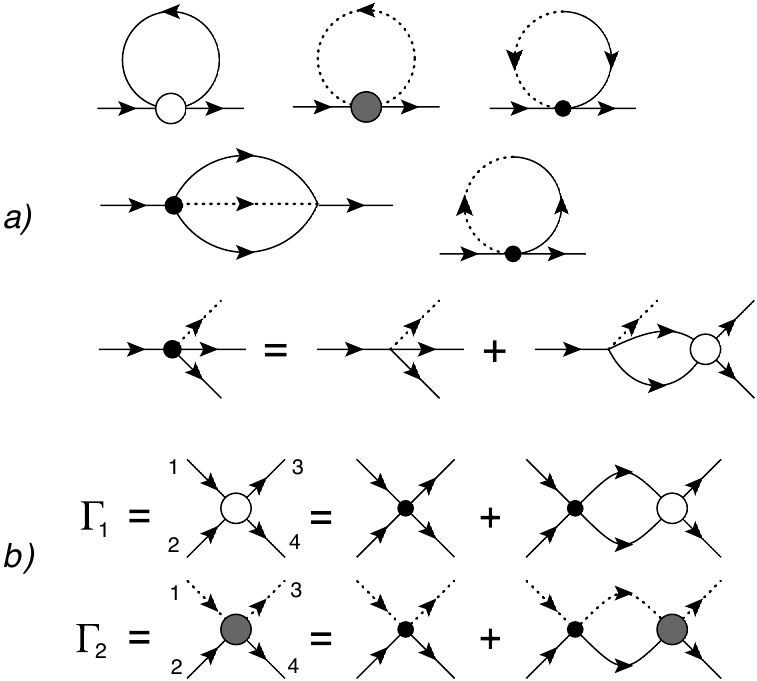}
\caption{(a) Diagrams for the normal self-energy part $\Sigma_p$ giving non-zero contributions of the second order in $J/D$. Solid and dashed lines stand for Green's functions $G(p)$ of $a$ and $b$ particles, respectively, defined by Eq.~\eqref{g}. Lines containing solid and dashed parts stand for anomalous Green's functions $F(p)$ defined by Eq.~\eqref{f}. Bare vertexes are defined by Eqs.~\eqref{aaaa} and \eqref{abbb}. Open and filled circles represent renormalized vertexes equations for which are presented in panel (b).}
\label{diag1}
\end{figure}

The elementary excitation spectrum $\e_{\pp}$ is defined by poles of Green's functions \eqref{g} and \eqref{f}:
\begin{eqnarray} 
\label{speceq}
 {\cal D}( \e_{\pp} , \pp ) = 0.
\end{eqnarray}
Let us consider diagrams for the normal self-energy part $\Sigma_p$ some of which are shown in Fig.~\ref{diag1}. If a diagram contains a contour that can be walked around while moving by arrows of functions $G_{0}(p)$, integrals over frequencies in such a diagram give zero. \cite{popov} In diagrams without such contours there are at least two vertexes \eqref{abbb} or at least one vertex \eqref{abbb} and one anomalous Green's functions $F(p)$. As the vertex \eqref{abbb} is of the order of $J/D$ and the numerator of $F(p)$ is $O(J/D)$ (see Eq.~\eqref{h2an}), the normal self-energy part is the value of the order of $(J/D)^{2}$. Thus, we obtain for $\Sigma_p$ in the first order in $J/D$
\begin{equation}
\label{sigma1}
 \Sig_p^{(1)} = 0.
\end{equation}

\begin{figure}
\includegraphics[scale = 1.0]{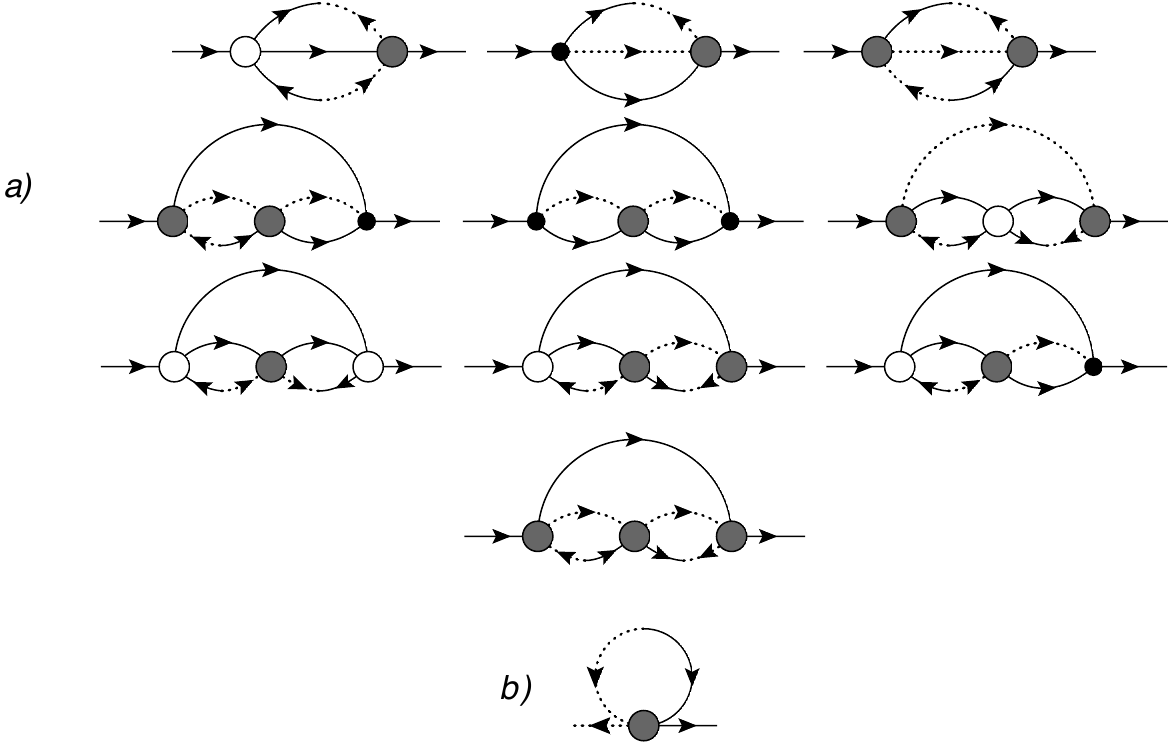}
\caption{(a) Diagrams for the normal self-energy part $\Sigma_{p}$ giving non-zero contributions of the third order in $J/D$. (b) The diagram for anomalous self-energy part $\Pi_{p}$ of the second order in $J/D$. Same notation as in Fig.~\ref{diag1}.}
\label{diag3}
\end{figure}

Let us turn to diagrams for the anomalous self-energy parts $\Pi_{p}$. One of them is shown in Fig.~\ref{diag3}(b). The contribution to $\Pi_{p}$ of the first order in $J/D$ is given by term \eqref{h2an}. Diagrams for $\Pi_{p}$ with one vertex contain sums like $\sum_{\kk} J_{\pp+\kk} = 0$ in the first order in $J/D$. Therefore such diagrams are at least of the second order in $J/D$. A consideration of other diagrams similar to that presented above for $\Sigma_p$ shows that they give contributions of the second order in $J/D$ and higher. Then we have in the first order in $J/D$
\begin{equation}
\label{pi1}
 \Pi^{(1)}_p = \frac{c_1^2}{2} J_\pp.
\end{equation}

It is convenient to rewrite Eq.~\eqref{speceq} using Eq.~\eqref{den} in the form
\beq 
\label{spec1}
 \e_{\pp}^{2} &=&  \left( \e_{1\pp} + \Sig_{(\e_{\pp}, \pp)} \right)^2 - \left|\Pi_{(\e_{\pp}, \pp)}\right|^2 + \left( \e_{\pp} - \e_{1\pp} - \Sig_{(\e_{\pp}, \pp)} \right) \left( \Sig_{(\e_{\pp}, \pp)} - \Sig_{(-\e_{\pp}, -\pp)}\right).
\eeq
It follows from the previous discussion that
\begin{eqnarray} 
\e_{\pp} - \e_{1\pp} - \Sig_{(\e_{\pp}, \pp)} = O((J/D)^2),\\
\Sig_{(\e_{\pp}, \pp)} - \Sig_{(-\e_{\pp}, -\pp)} = O((J/D)^2). \non
\end{eqnarray}
Bearing in mind these equations and that our aim is to find the spectrum in the third order in $J/D$ we can write Eq.~\eqref{spec1} in the form
\beq
\label{spec2}
\e_{\pp}^{2} = \left( \e_{1\pp} + \Sig_{(\e_{\pp}, \pp)} \right)^{2}- \left|\Pi_{(\e_{\pp}, \pp)}\right|^2 + O((J/D)^4).
\eeq
As follows from this equation, the first corrections to $\e_{1\pp}$ are of the second order in $J/D$. Taking into account that the first corrections to self-energy parts are also of the second order in $J/D$, we can replace $(\e_{\pp}, \pp)$ by $(\e_{1\pp}, \pp)$ in Eq.~\eqref{spec2} and write down the final formula for the spectrum calculation up to the third order in $J/D$
\begin{eqnarray}
\label{spec3} 
 \e_\pp^2 &=& \left( \e_{1\pp} + \Sig_{(\e_{1\pp},\pp)} \right)^2 - \left|\Pi_{(\e_{1\pp}, \pp)}\right|^2. \end{eqnarray}

Let us find the spectrum in the second order in $J/D$. As is clear from Eq.~\eqref{spec3}, one has to use the anomalous self-energy part in the first order in $J/D$ for which we obtain Eq.~\eqref{pi1}. The normal self-energy part has to be found in the second order in $J/D$ in which diagrams presented in Fig.~\ref{diag1}(a) should be taken into account. In order to calculate these diagrams it is necessary to find vertexes denoted by open and filled circles in the zeroth order. As follows from the diagram analysis presented above, the zeroth and the first order contributions to the vertexes are presented by a series of ladder diagrams only (see Fig.~\ref{diag1}(b)). Diagrammatic equations in Fig.~\ref{diag1}(b) have the following explicit form in the zeroth order in $J/D$:
\begin{subequations}
\label{vert0}
\begin{eqnarray}
 \Gamma_1^{(0)} &=& D - 2 \frac{ \Gamma_1^{(0)} D }{ 2D - \Omega}, \\
 \Gamma_2^{(0)} &=& U - \frac{ \Gamma_2^{(0)} U }{ 2D - \Omega },
\end{eqnarray}
\end{subequations}
where $\Omega$ is the sum of incoming lines frequencies. The solution of Eq.~\eqref{vert0} at $ U \to +\infty $ has a simple form
\begin{subequations}
\label{vert1}
\begin{eqnarray}
 \Gamma_1^{(0)} &=& \frac{ D \left( 2D - \Omega \right)}{ 4D - \Omega }, \\
 \Gamma_2^{(0)} &=& 2D - \Omega.
\end{eqnarray}
\end{subequations}

After substitution of Eqs.~\eqref{vert1} to diagrams for self-energy parts and integration over frequencies $\Omega$ gives a difference of the kind $\epsilon_{1{\bf k}_1}-\epsilon_{1{\bf k}_2}$ with some momenta ${\bf k}_{1,2}$. Then $\Omega$ corresponds to values of the order of $J/D$ in diagrams for self-energy parts and it can be neglected in the vertex calculation in the zeroth order in $J/D$. As a result we obtain for vertexes from Eqs.~\eqref{vert1}
\begin{subequations}
\label{ver0res}
\begin{eqnarray}
 \Gamma_1^{(0)} &=& D/2, \\
 \Gamma_2^{(0)} &=& 2D.
\end{eqnarray}
\end{subequations}

One obtains for the normal self-energy part in the second order in $J/D$ as a result of calculation of diagrams shown in Fig.~\ref{diag1}(a) taking into account Eqs.~\eqref{ver0res}
\begin{equation}
\label{sigmas2}
\Sig^{(2)}_{( \e_{1\pp}, \pp )} =  \frac{2c_1^4 + 4c_1 c_2 - c_1^2 c_2^2}{8D} \frac1N\sum_{\kk} J^{2}_{\kk}. 
\end{equation}
Using Eqs.~\eqref{sigmas2}, \eqref{spec3} and \eqref{pi1} we have for the spectrum in the second order in $J/D$ (cf.\ Eq.~\eqref{spec0})
\begin{equation}
\label{specs2}
\e_{2\pp} = \e_{1\pp} + \Sig^{(2)}_{( \e_{1\pp}, \pp )} - \frac{1}{2\e_{0\pp}}\left|\Pi^{(1)}_{(\e_{1\pp}, \pp)}\right|^2 =   D + \frac{c_1^2}{2} J_\pp + \frac{2c_1^4 + 4c_1 c_2 - c_1^2 c_2^2}{8D} \frac1N\sum_{\kk} J^{2}_{\kk} - \frac{ c_1^4 } {8} \frac{J_\pp^2} {D}. 
\end{equation}

In order to calculate the spectrum in the third order in $J/D$ one has to find the normal and anomalous self-energy parts in the third and the second order, respectively. To find $\Sigma_{p}$ in the third order we have to take into account diagrams shown in Fig.~\ref{diag3}(a). Besides, one has to consider also diagrams presented in Fig.~\ref{diag1}(a) taking into account the first order corrections to the vertexes for which we have after solving equations in Fig.~\ref{diag1}(b)
\begin{eqnarray}
\Gamma_1^{(1)} &=& \frac12 D - \frac18 \Omega + \frac12 J_{3-1} - \frac{c_1 c_2}{4} (J_1 + J_3), \\
\label{gamma2}
\Gamma_2^{(1)} &=& 2D - \Omega - J_{3-1}.
\end{eqnarray}
The only diagram of the second order in $J/D$ for the anomalous self-energy part is shown in Fig.~\ref{diag3}(b). It should be found using Eq.~\eqref{gamma2}. As a result of straightforward calculation we obtain
\begin{eqnarray}
\label{sig3}
	\Sigma^{(3)}_{( \e_{1\pp}, \pp )} &=& \frac{c_1^2(c_1 + c_2)^2}{16D^2}\frac{1}{N^2} \sum_{\kk,\qq} J_\qq J_\kk J_{\qq-\kk} - \frac{ 2c_1^4 + 4 c_1 c_2 - c_1^2 c_2^2 }{8D^2} J_\pp \frac{1}{N} \sum_\kk J^2_\kk 
\non\\
&&{}
+
\frac{c_1^4 + 3 c_1^2 c_2^2 - 2 c_1^3 c_2 + c_1 c_2 }{8D^2} \frac{1}{N^2} \sum_{\kk,\qq} J_\pp J_\kk J_{\qq-\kk + \pp},\\
\label{pi2}
\Pi^{(2)}_{( \e_{1\pp}, \pp )} &=& \frac{c_1^2}{4D} \frac{1}{N} \sum_\kk J_\kk J_{\kk-\pp} + \frac{c_{1}^{4}}{4D} \frac1N \sum_{\kk} J^{2}_{\kk}.
\end{eqnarray}
Using Eqs.~\eqref{spec3}, \eqref{sigma1}, \eqref{pi1}, \eqref{sigmas2}, \eqref{pi2} and \eqref{sig3}, one has for the spectrum in the third order in $J/D$
\begin{eqnarray}
\label{specs3}
\e_{3\pp} &=& D + \frac{c_1^2}{2} J_\pp + \frac{ 2c_1^4 + 4 c_1 c_2 - c_1^2 c_2^2 }{8D} \frac{1}{N} \sum_\kk J^2_\kk - \frac{c_{1}^{4}}{8D} \frac1N J^{2}_{\pp} \non\\
&&{}+ \frac{c_1^2(c_1 + c_2)^2}{16D^2}\frac{1}{N^2} \sum_{\kk,\qq} J_\qq J_\kk J_{\qq-\kk} - \frac{ 4c_1^4 + 4 c_1 c_2 - c_1^2 c_2^2 }{8D^2} J_\pp \frac{1}{N} \sum_\kk J^2_\kk \non \\
&&{}+ \frac{c_1^4 + 3 c_1^2 c_2^2 - 2 c_1^3 c_2 + c_1 c_2 }{8D^2} \frac{1}{N^2} \sum_{\kk,\qq} J_\pp J_\kk J_{\qq-\kk + \pp} - \frac{c_1^4}{8D^2} J_\pp \frac{1}{N} \sum_\kk J_\kk J_{\kk-\pp} + \frac{c_1^6}{16D^2} J_\pp^3.
\end{eqnarray}

The spectrum $\e_{3\pp}$ has a minimum in some point $\pp=\pp_0$, in which it is separated from the ground state by a gap. For example, this minimum is located at $\pp$ equal to the antiferromagnetic vector $\pp_0$ in the simple square or the simple cubic lattices with antiferromagnetic exchange interaction between only nearest neighbors ($\pp_0=(\pi,\pi)$ and $(\pi,\pi,\pi)$, respectively). The gap decreases with decreasing $D$ and it vanishes at QCP $D=D_{c}$.	Upon further decreasing of $D$ a ``condensation'' takes place of elementary excitations with momentum $\ppp$ which corresponds to appearance of long-range magnetic order. We show below by particular examples that at $D\agt D_c$ series in $J/D$ for some quantities converge slowly. It is a manifestation of strong fluctuations near QCP.

It is worth noting that $c_1,c_2\sim S$ at $S\gg1$ (see Eqs.~\eqref{c1} and \eqref{c2}). Therefore, as follows from Eq.~\eqref{specs3}, the expansion parameter is actually $S^2J/D$ (not $J/D$) at $S\gg1$. It means, in particular, that $D_c\sim S^2J$ in the case of antiferromagnetic exchange interaction between nearest neighbor spins on the simple square or the simple cubic lattices. This conclusion is consistent with the result of the spin wave analysis carried out in antiferromagnetic phase (see, e.g., Ref.~\cite{prev}). Thus, a very strong anisotropy or a very small exchange are required for the paramagnetic phase stability at $S\gg1$ and the paramagnetic phase is absent in the classical spin limit at any finite $J$ and $D$ as it must be.

\subsection{Ground state energy}

\begin{figure}
 \includegraphics[scale = 1.0]{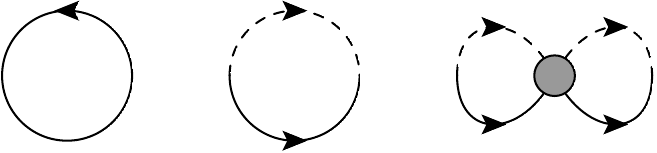}
 \caption{ Diagrams giving non-zero contributions of the second and the third order in $J/D$ to the ground state energy. Same notation as in Fig.~\ref{diag1}.}
 \label{gsfig}
\end{figure}

It is useful to calculate the ground state energy for the sake of comparison with numerical results. Diagrams giving non-zero contributions of the second and the third order in $J/D$ are shown in Fig.~\ref{gsfig}. The straightforward calculation of these diagrams leads to the following expression for the ground state energy:
\begin{equation}
\label{gs}
E_{gs} = -\frac{ c_1^4 }{ 8D } \frac1N \sum_\pp J^2_\pp - \frac{ c_1^4 }{ 16D^2 } \frac{1}{N^2} \sum_{\pp,\kk} J_\pp J_\kk J_{\pp+\kk}.
\end{equation}
Notice that the first non-zero term in $E_{gs}$ is of the second order in $J/D$.

\section{Discussion and comparison with previous results and experiment}
\label{disc}

Note that we do not specify the type of exchange interaction $J_{ij}$ in Eq.~\eqref{ham} and the lattice type and dimension while deriving Eq.~\eqref{specs3}. Then Eq.~\eqref{specs3} is applicable, in particular, for the spin chain with $S=1$ and with exchange interaction between nearest neighbors only. The spectrum in this special case was calculated before in the third order in $J/D$ in Ref.~\cite{4} using the regular perturbation theory. It is easy to verify that the result of Ref.~\cite{4} coincides with Eq.~\eqref{specs3} in this case.

\subsection{Other spin representations for $S=1$}
\label{secSEqualOne}

In the particular case of $S=1$ a number of simpler spin representations can be introduced. In one of them $S^{z}_{i}$ is given by Eq.~\eqref{sz} and $S^+_i$ has the form (cf.\ Eq.~\eqref{s+})
\begin{eqnarray}
\label{s+simple}  
S^+_i = \sqrt{2} \left( \bkr_i + a_i \right). 
\end{eqnarray}
In contrast to Eqs.~\eqref{sz}--\eqref{s+} the introduction of term \eqref{u2} into the Hamiltonian is necessary now because Eq.~\eqref{s+simple} has non-zero matrix elements between states from physical and unphysical subspaces. Representation \eqref{s+simple} is actually introduced in Ref.~\cite{kolezh}, where spin chains in magnetic field are discussed. We have made calculations of the spectrum and the ground state energy using this representation and recovered Eqs.~\eqref{specs3} and \eqref{gs} at $S=1$.

It should be noted that at $S=1$ one can make calculations without taking into account the constraint term \eqref{u2} by adding to Eq.~\eqref{s+simple} projector operators $1 - n_{a,i} - n_{b,i}$ as follows: 
\begin{equation}
\label{s+no}
S^{+}_i = \sqrt2 \left[  (1 - n_{a,i} - n_{b,i}) b_i  +  \akr_i  (1 - n_{a,i} - n_{b,i}) \right]. 
\end{equation}
In contrast to Eq.~\eqref{s+simple} this representation has zero matrix elements between states from physical and unphysical subspaces and there is no need to take into account Eq.~\eqref{u2} now. We have carried out calculations with Eq.~\eqref{s+no} and recovered Eqs.~\eqref{specs3} and \eqref{gs} at $S=1$ once again.

The following spin representation for $S=1$ is introduced in Ref.~\cite{3}:
\beq
\label{s+sq}
S^{+}_i = \sqrt2 \left[  \sqrt{1 - n_{a,i} - n_{b,i}}\; b_i  +  \akr_i  \sqrt{1 - n_{a,i} - n_{b,i}} \right],
\eeq
which is equivalent to our one \eqref{s+no} on the physical subspace and which, therefore, should lead to the same results. But authors of Ref.~\cite{3} do not take advantage of the opportunity to find physical quantities in the form of series in terms of powers of $J/D$ which such representations provide. They expand roots in Eq.~\eqref{s+sq} and analyze the spectrum of the Hamiltonian in the harmonic approximation. They also find renormalization of this spectrum by taking into account the simplest diagrams and carrying out some self-consistent calculations for the square lattice and antiferromagnetic exchange. We compare below results obtained in Ref.~\cite{3} with our ones.

\subsection{ Other approaches to the spectrum calculation }

After substitution of the self-energy parts \eqref{sigma1} and \eqref{pi1} obtained in the first order in $J/D$ to the general expression \eqref{spec1} for the spectrum we recover Lindgard's formula \cite{lind1,lind2,11}
\begin{eqnarray}
\label{eqLin}	
 \e_{L\pp} = \sqrt{ D \left[ D + S(S+1) J_\pp \right] },
\end{eqnarray}
which was found in Refs.~\cite{lind1,lind2} within the random phase approximation. But as it is clear from the above discussion, Eq.~\eqref{eqLin} is correct only in the first order in $J/D$. The second and the third order terms in $J/D$ in Eq.~\eqref{eqLin} differ significantly from those in Eq.~\eqref{specs3}. Case study that is done below shows that Eq.~\eqref{eqLin} works quite badly when $J/D$ is not very small.

A "generalized spin-wave approach" (GSWA) is used in Refs.~\cite{13,chern} for the spectrum consideration in DTN. In the framework of this approach in which fluctuations are taken into account in a mean field fashion the following expression is obtained:
\begin{equation}
\label{gswa}
\e_\pp^{gswa}=\sqrt{\mu(\mu+2s^2 J_\pp)},
\end{equation}
where parameters $s$ and $\mu$ are determined as a result of self-consistent calculations using equations
\begin{eqnarray}
D &=& \mu\left( 1+\frac1N\sum_\pp\frac{J_\pp}{\e_\pp^{gswa}} \right),\\
\label{gswas}
s^2 &=& 2 - \frac1N\sum_\pp\frac{\mu+s^2J_\pp}{\e_\pp^{gswa}}. 
\end{eqnarray}
It is seen that Eq.~\eqref{gswa} is a modification of Eq.~\eqref{eqLin} at $S=1$: $D$ and $J_\pp$ are renormalized. We show below by case study that Eqs.~\eqref{gswa}--\eqref{gswas} work much better than Eq.~\eqref{eqLin} but Eq.~\eqref{specs3} proves to be more precise and convenient because the numerical solving of integral equations and numerical calculations are not required.

It should be noted that an approach very similar to that leading to Eqs.~\eqref{gswa}--\eqref{gswas} is proposed in Refs.~\cite{1,wang5}. However the resultant equations in Refs.~\cite{1,wang5} are more cumbersome than Eqs.~\eqref{gswa}--\eqref{gswas} whereas they work slightly worse than Eqs.~\eqref{gswa}--\eqref{gswas} as our comparison shows with the numerical results of Refs.~\cite{2,3}. That is why we do not consider here in detail results of Refs.~\cite{1,wang5}.

\subsection{ Comparison with numerical results }
\label{secMC}

\begin{figure}
 \includegraphics[scale=0.6]{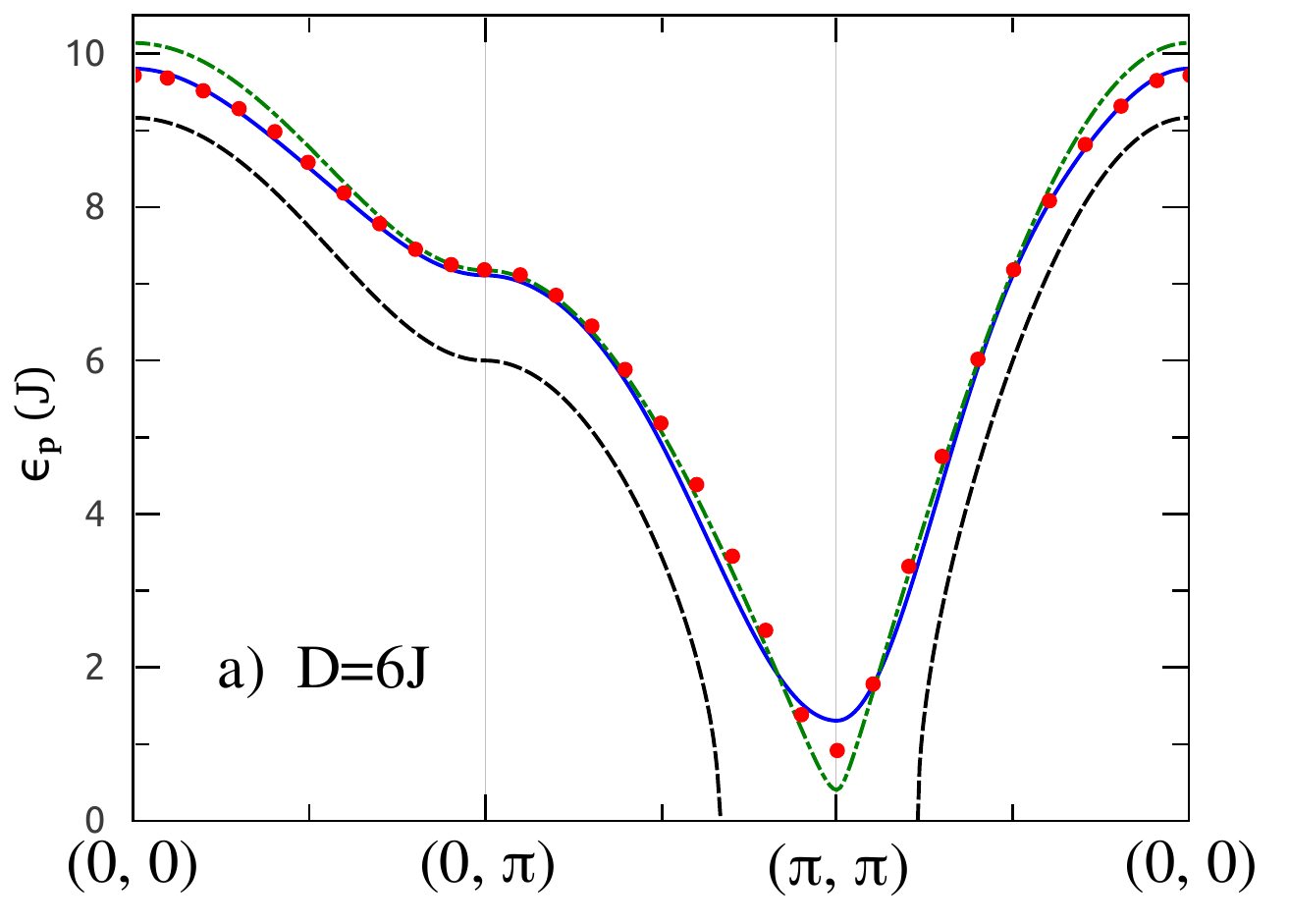}
 \includegraphics[scale=0.6]{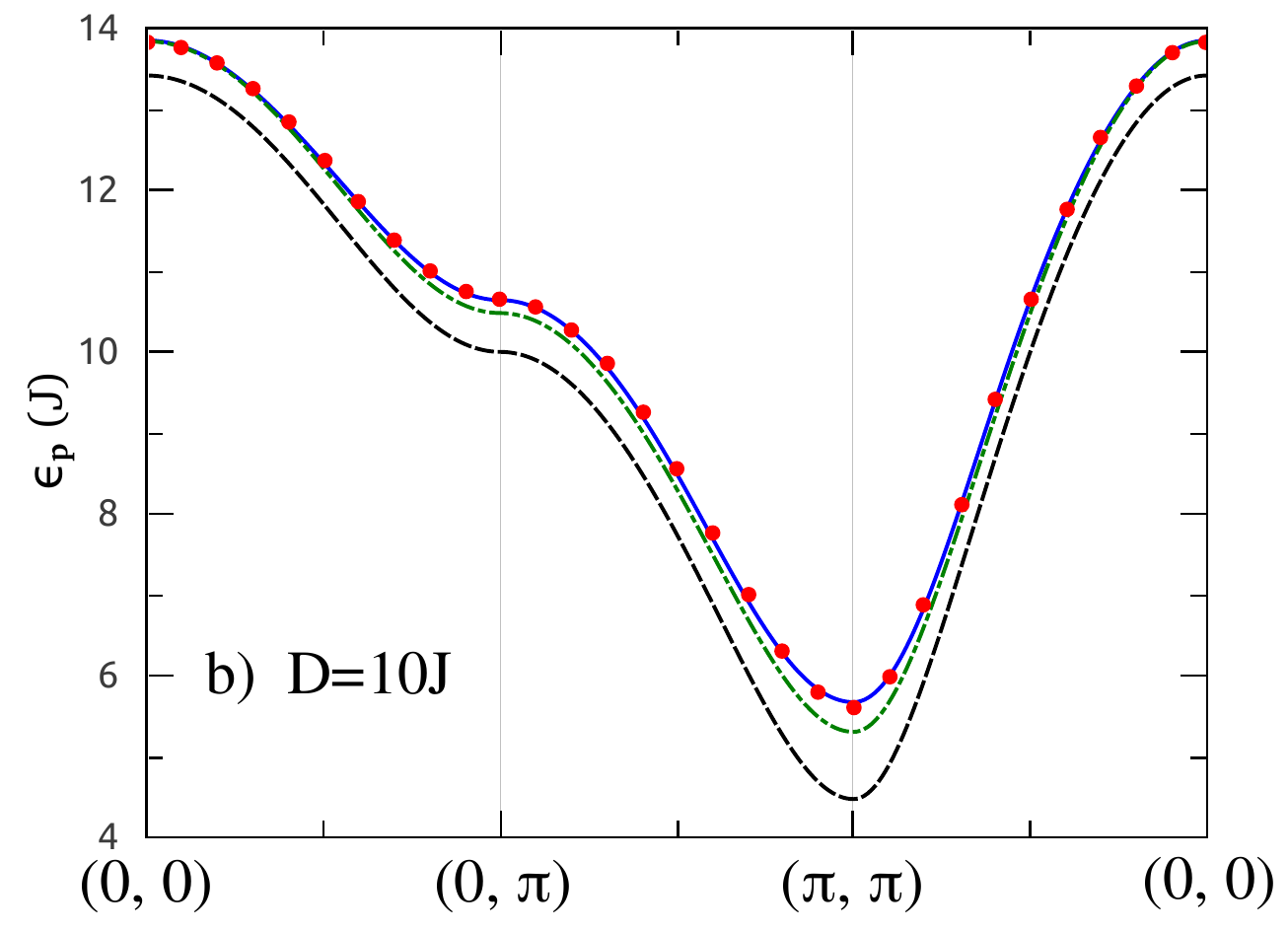}
 \caption{(Color online.) Elementary excitation spectrum $\e_\pp$ of the model \eqref{ham} on the square lattice with $S=1$ and antiferromagnetic exchange $J>0$ between nearest neighbors for (a) $D=6J$ and (b) $D=10J$. Monte-Carlo data of Refs.~\cite{2,3} are shown by points. Dash, dash-dotted and solid lines are drawn using Lindgard's formula \eqref{eqLin}, GSWA formula \eqref{gswa} and Eq.~\eqref{specs3} (the present paper result), respectively. The deviation of the spectrum \eqref{specs3} from the numerical data at $D=6J$ in the close vicinity of the minimum at $(\pi,\pi)$ is a consequence of the proximity to the QCP $D=D_c^{2D}\approx5.65$ separating the paramagnetic and the antiferromagnetic phases (see the main text).}
 \label{figMC}
\end{figure}

The elementary excitation spectrum is found in Ref.~\cite{2} by Monte-Carlo calculations for the square lattice and $S=1$. The exchange interaction is taken in Ref.~\cite{2} to be positive (antiferromagnetic) and equal for all nearest neighbors. Results of Refs.~\cite{2,3} together with the spectrum calculated using Eqs.~\eqref{specs3}, \eqref{gswa}--\eqref{gswas} and \eqref{eqLin} are shown in Fig.~\ref{figMC} for $D=6J$ and $D=10J$. It is seen from  Fig.~\ref{figMC} that Eq.~\eqref{specs3} works well in the whole Brillouin zone when $D$ is not very close to the critical value $D_c^{2D}\approx5.65$ found numerically in Refs.~\cite{haas,2,wong}. The phase with the long-range magnetic order (antiferromagnetic phase) is stable at $D<D_c^{2D}$. Notice also that even at $D\agt D^{2D}_c$ Eq.~\eqref{specs3} describes well the spectrum of short-wavelength elementary excitations.

\begin{figure}
 \includegraphics[scale=0.7]{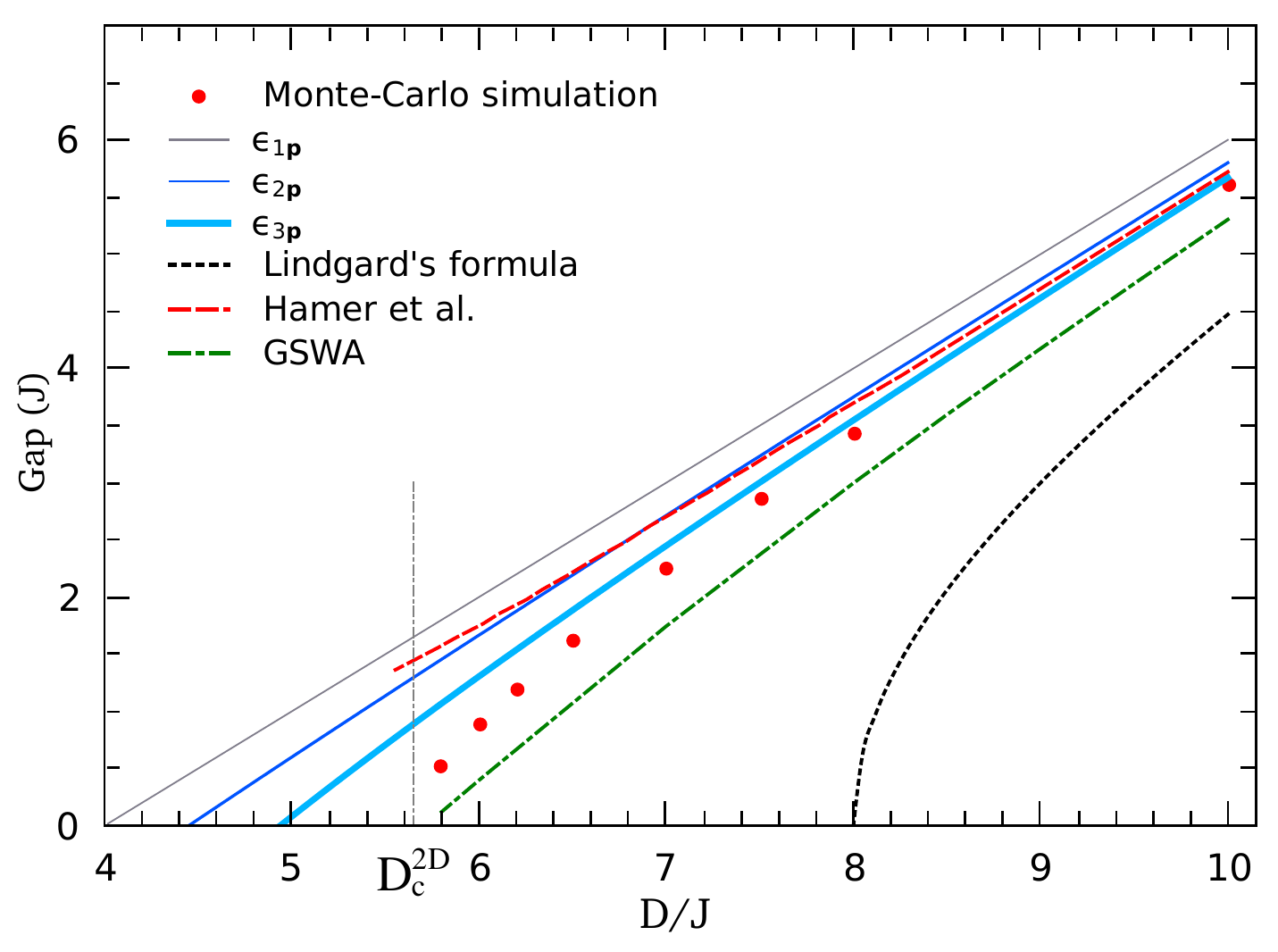}
 \caption{ 
(Color online.) The spectrum gap at $\pp=(\pi,\pi)$ is presented as a function of $D$ in the model \eqref{ham} with $S=1$ and antiferromagnetic exchange interaction $J$ between only nearest neighbors. Monte-Carlo data of Ref.~\cite{2,3} are shown by points, results by Hamer et al. \cite{3} are represented by dash line, dash-dotted and dotted lines are drawn using GSWA formula \eqref{gswa} and Lindgard's formula \eqref{eqLin}, respectively. Spectra $\e_{1\pp}$, $\e_{2\pp}$ and $\e_{3\pp}$ are also shown, which are found in the first, the second and the third orders in $J/D$ and which are given by Eqs.~\eqref{spec0}, \eqref{specs2} and \eqref{specs3}, respectively. The critical value of the anisotropy $D_c^{2D}\approx5.65$ \cite{haas,2} is marked, below which the phase with the antiferromagnetic long-range magnetic order is stable.}
 \label{figgap}
\end{figure}

The noticeable deviation of the long-wavelength excitation spectrum given by Eq.~\eqref{specs3} (with momenta ${\bf p}\approx\pp_0=(\pi,\pi)$) at $D\agt D^{2D}_c$ is a result of strong fluctuations near QCP which manifest themselves in a bad convergence of the series in terms of powers of $J/D$. It is illustrated by Fig.~\ref{figgap} which shows the dependence of the spectrum gap at $\pp=\ppp$ on $D$. It is seen that the second and the third order corrections to the gap are approximately equal to each other at $D\agt D^{2D}_c$ and the value of $D_{c}$ found from the relation $\e_{3\pp_0}=0$ is 12\% smaller than the value of $D_c^{2D}\approx5.65$ obtained numerically in Refs.~\cite{haas,2,wong}.

\begin{figure}
 \includegraphics[scale=0.5]{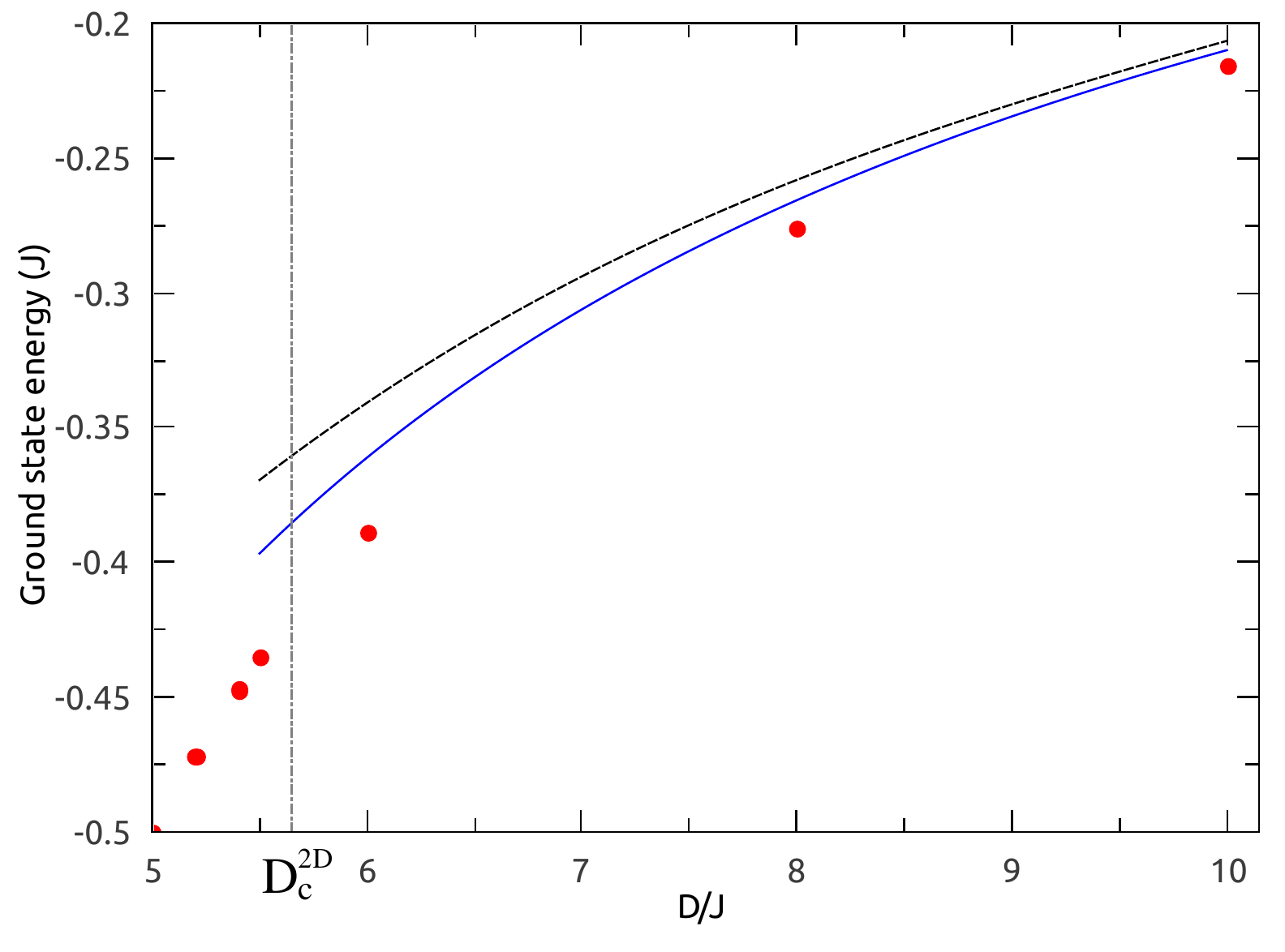}
 \caption{(Color online.) The ground state energy of the model \eqref{ham} on the square lattice with $S=1$ and antiferromagnetic exchange between only nearest neighbors. Monte-Carlo data of Refs.~\cite{2,3} are shown by points, results by Hamer et al. \cite{3} are presented by dash line, the solid line is drawn using Eq.~\eqref{gs}, which is obtained in the third order in $J/D$.}
 \label{figgs}
\end{figure}

The result of the ground state energy calculation is shown in Fig.~\ref{figgs}, which demonstrates that Eq.~\eqref{gs} works well when $D$ is not very close to $D^{2D}_c$ and that series for $E_{gs}$ converges slowly at $D\agt  D^{2D}_c$.

Comparison of the results obtained within different approaches and presented in Figs.~\ref{figMC}--\ref{figgs} shows that formulas derived above using the method suggested in the present paper, on the whole, work better. At the same time it should be noted that despite its simplicity GSWA gives quite precise results.

\subsection{Application to $\rm NiCl_2$-$\rm 4SC(NH_2)_2$}

At the present time the most extensively studied compound described by the model \eqref{ham} and having the paramagnetic ground state is $\rm NiCl_2$--$\rm 4SC(NH_2)_2$ which is known as DTN \cite{12,13,14,15,16,17,chern,add8}. The magnetic subsystem of DTN consists of $\rm Ni$ ions with $S=1$ and the Lande factor $g=2.26$. Magnetic ions form a body-centered tetragonal lattice which can be viewed as two interpenetrating tetragonal sublattices. The exchange interaction between spins inside one sublattice is antiferromagnetic and strongly anisotropic: the exchange constant along the tetragonal hard axis ($z$ axis) is much larger than those along $x$ and $y$ axes. That is why DTN is considered as a quasi-1D compound. Hamiltonian \eqref{ham} with the following set of parameters is used for interpretation of the majority of experimental data:
\beq
 \label{olddtnpar}
D &=& 8.9\, \rm{K}, \non \\
J_z &=& 2.2\, \rm{K},\\
J_{xy} &=& 0.18\, \rm{K}, \non
\eeq
where $J_{z}$ is the exchange constant along the chains and $J_{xy}$ is the exchange coupling constants between chains inside one tetragonal sublattice. Interaction between tetragonal sublattices is supposed to be negligibly small \cite{13} in the most of considerations.

DTN behavior attracts special attention near two QCPs in magnetic field $H$ applied along the hard $z$ axis. The first QCP $H=H_{c1}$ separates the paramagnetic and the canted antiferromagnetic phases and the second one separates the canted antiferromagnetic phase and the ferromagnetic one in which all spins are parallel to the field. Values of these critical fields are defined by the following {\it exact} relations (see, e.g., Refs.~\cite{13,chern}):
\begin{eqnarray}
\label{hc1}
 H_{c1} &=& \e_{\pp_0}, \\
\label{hc2}
 H_{c2} &=& 2 S J_{\bf 0} + V_{\bf0}+ D (2 S - 1),
\end{eqnarray}
where $\e_{\pp}$ is the spectrum at $H=0$ and $V$ is the exchange coupling between sublattices considered below which has been neglected so far. The following values of the critical fields are obtained experimentally in Ref.~\cite{16}:
\footnote{It should be noted that in some other papers (see, e.g., Refs.~\cite{12,14}) other values of the critical fields are reported: $H_{c1}=2.1$~T and $H_{c2}=12.6$~T. In the present paper we use values \eqref{hcdtn} because they were measured at extremely small temperature of 1~mK. Besides, our fitting of the neutron data of Ref.~\cite{13} by varying exchange coupling constants and $D$ while keeping the critical fields fixed gives better result with values \eqref{hcdtn}.
}
\begin{eqnarray}
\label{hcdtn}
\begin{array}{ll}
 H_{c1}^{DTN} &= 2.05\,{\rm T}, \\
 H_{c2}^{DTN} &= 12.175\, {\rm T}.
\end{array}
\end{eqnarray}

\begin{figure}
 \includegraphics[scale=0.55]{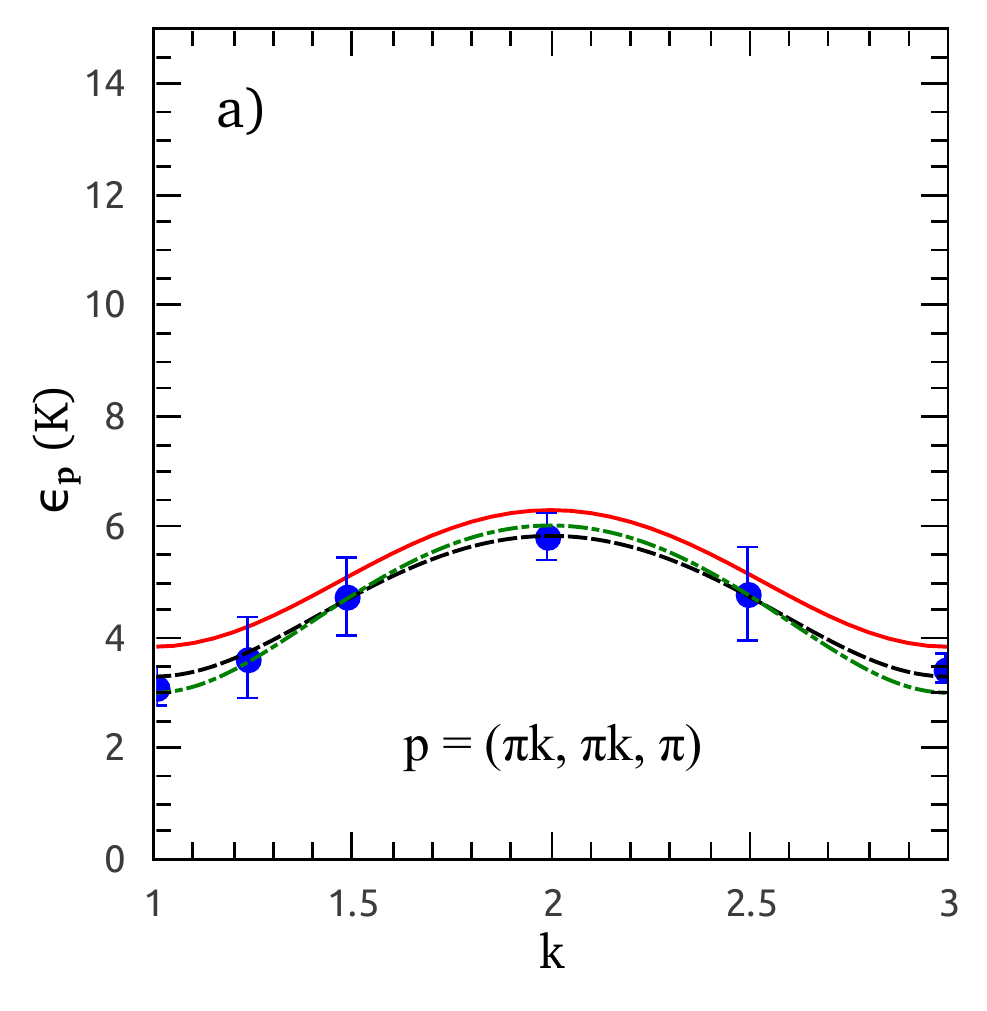}
 \includegraphics[scale=0.55]{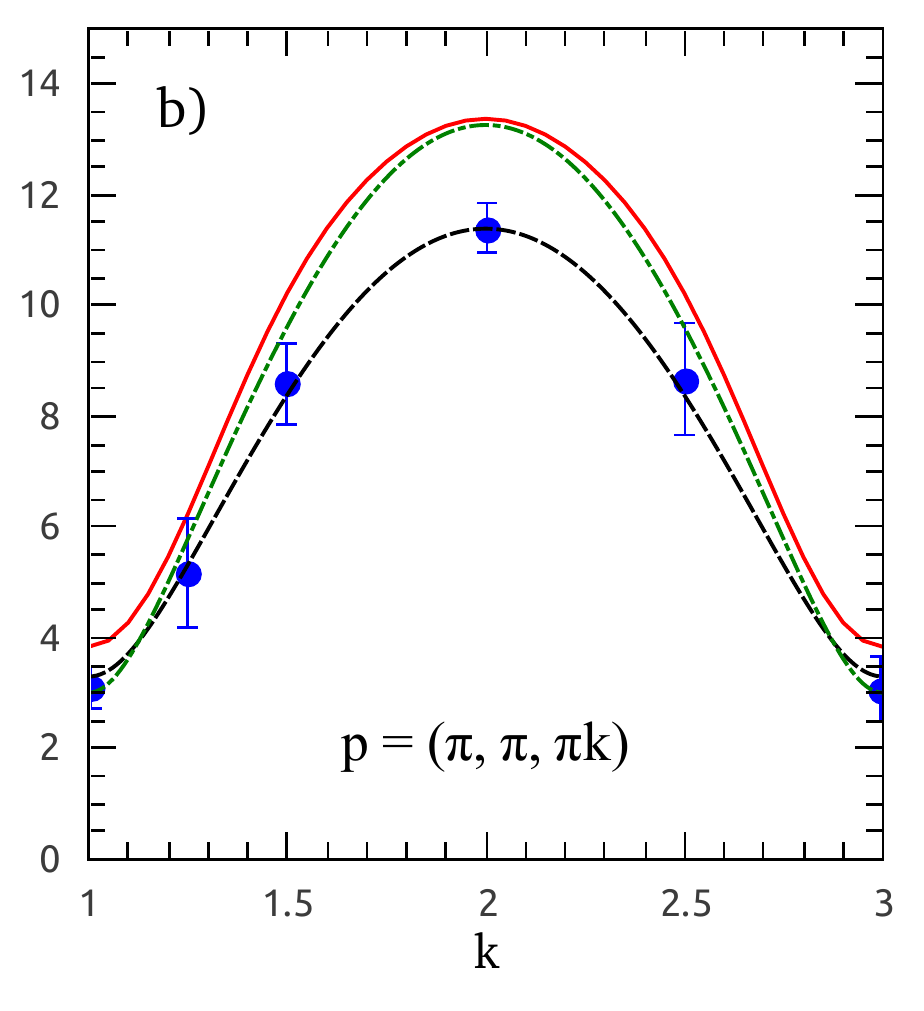}
 \includegraphics[scale=0.55]{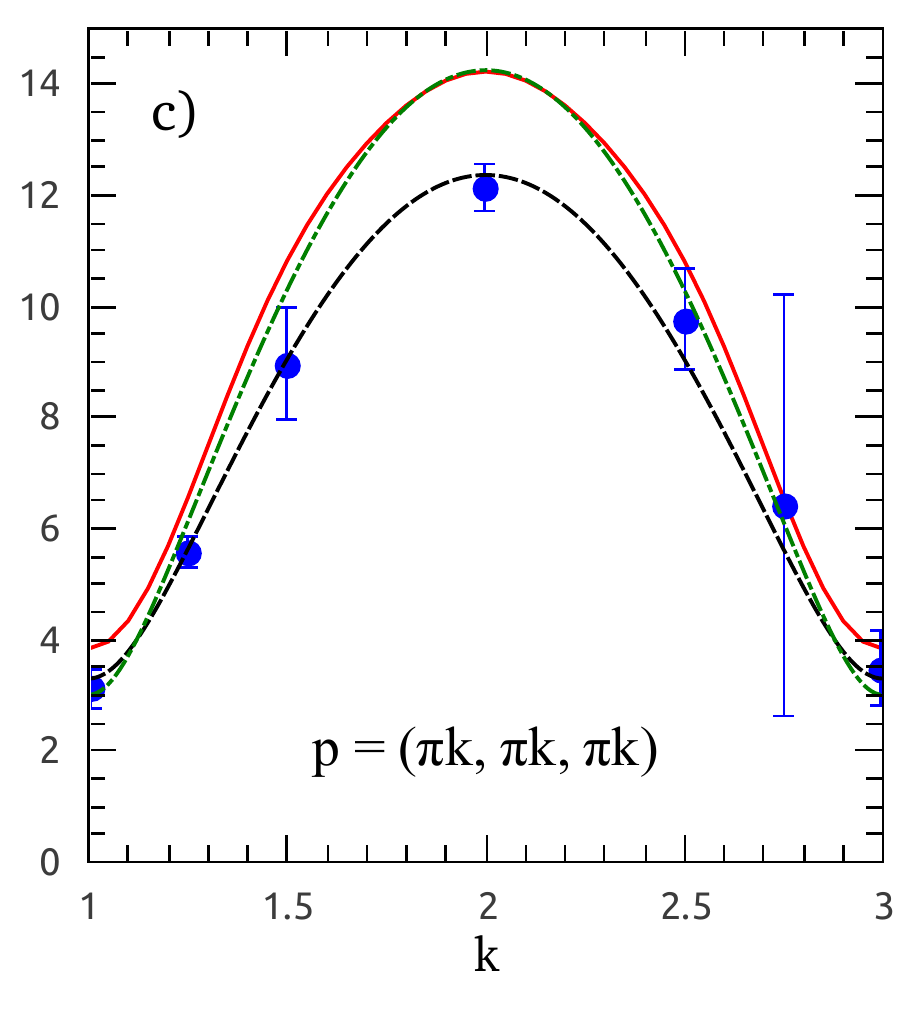}
 \includegraphics[scale=0.55]{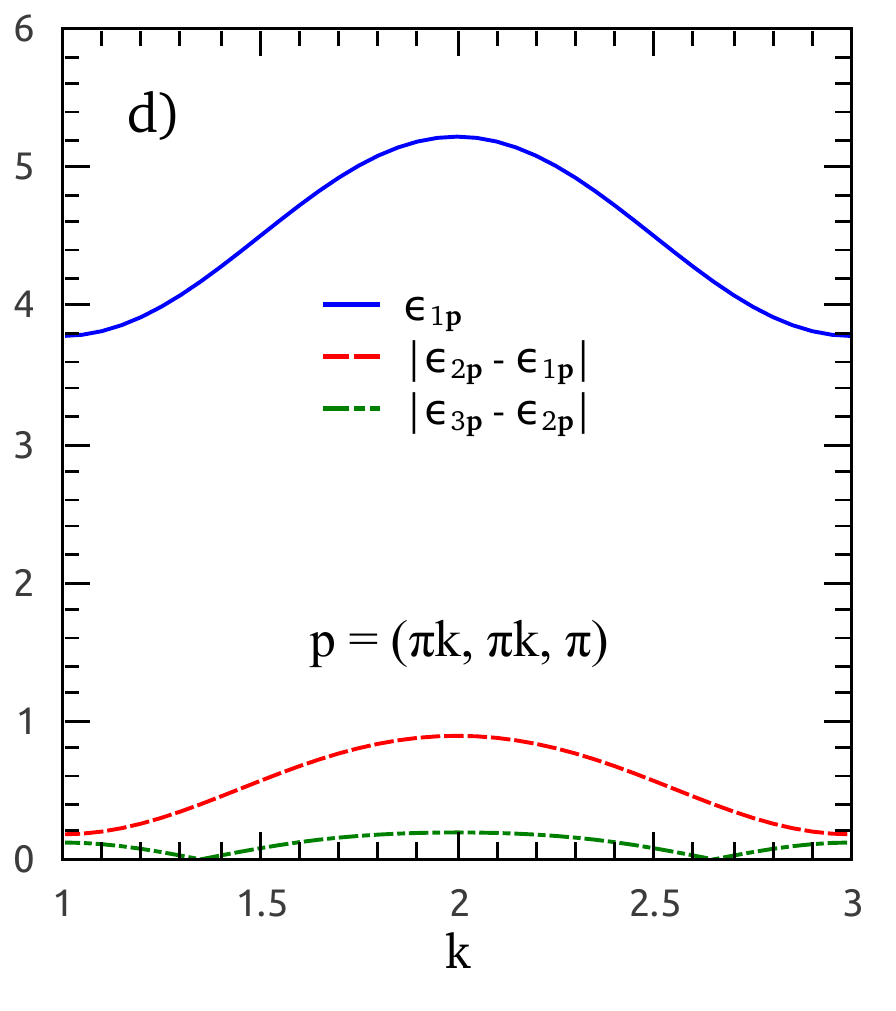}
 \includegraphics[scale=0.55]{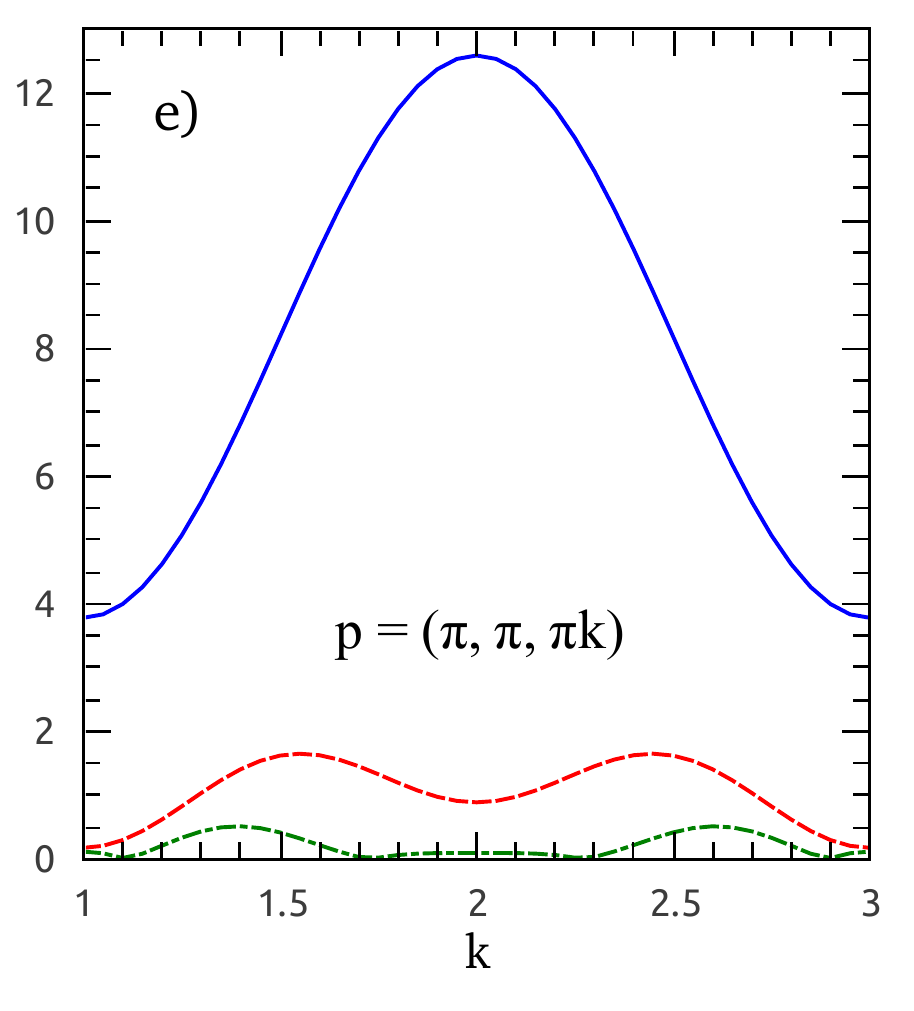}
 \includegraphics[scale=0.55]{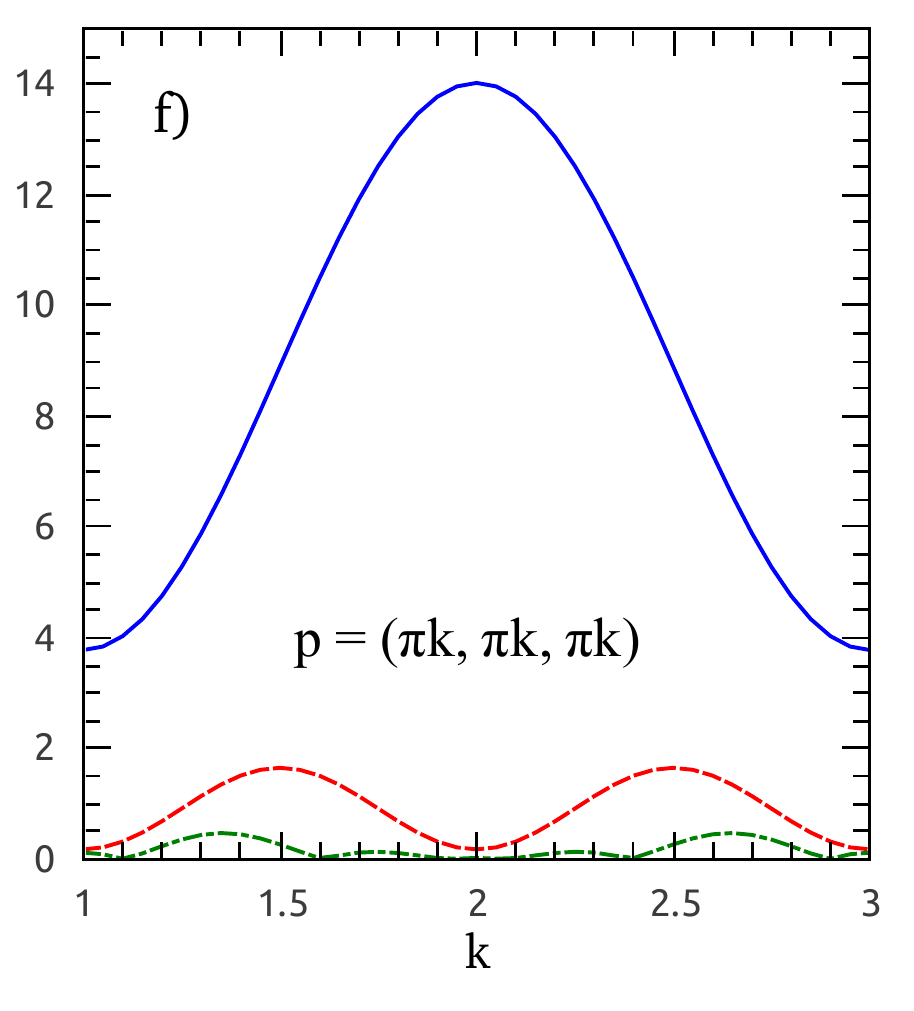}
 \caption{(Color online.) (a)--(c) Elementary excitation spectrum of DTN along three directions in the Brillouin zone at zero field. The data of the neutron experiment \cite{13} at $T=80$~mK are shown by points, dash and dash-dotted lines are drawn using Eqs.~\eqref{gswa}--\eqref{gswas} with parameters $D=8.12$~K, $J_{xy}=0.17$~K and $J_z=1.74$~K (which are proposed in Ref.~\cite{13}) and with the conventional set of parameters \eqref{olddtnpar}, respectively, solid lines are drawn using Eq.~\eqref{specs3} with the conventional parameters \eqref{olddtnpar}. (d)--(f) The spectrum of the first order in $J/D$ given by Eq.~\eqref{spec0} and the second and the third order corrections in $J/D$ to the spectrum found using Eqs.~\eqref{specs2} and \eqref{specs3} and the conventional set of parameters \eqref{olddtnpar}.}
 \label{figolddtn}
\end{figure}

In the present paper we focus on analysis of the DTN elementary excitation spectrum at zero magnetic field, which is observed in neutron experiment \cite{13} and shown in Fig.~\ref{figolddtn}(a)--(c). Its current theoretical interpretation looks inconsistent. Eqs.~\eqref{gswa}--\eqref{gswas} are used for the spectrum analysis in Refs.~\cite{13,chern}. As is shown in Ref.~\cite{13}, Eqs.~\eqref{gswa}--\eqref{gswas} describe DTN spectrum very well with parameters $D=8.12$~K, $J_{xy}=0.17$~K and $J_z=1.74$~K which differ from those used in the literature now \eqref{olddtnpar} (see the dash line in Fig.~\ref{figolddtn}(a)--(c)). However in the subsequent paper \cite{14} these parameters were declined because the value of the critical field $H_{c2}$ found using Eq.~\eqref{hc2} with these parameters differs significantly from the experimentally obtained value. The set of parameters \eqref{olddtnpar} is proposed in Ref.~\cite{14}, which has been used up to the present. In particular, in the recent paper \cite{chern} Eqs.~\eqref{gswa}--\eqref{gswas} are used with the conventional parameters \eqref{olddtnpar} for the spectrum analysis in the paramagnetic phase near the antiferromagnetic vector $\pp_0$. However, as it is seen from Fig.~\ref{figolddtn}(a)--(c) (the dash-dotted line), GSWA with this set of parameters describes unsatisfactorily the spectrum of short-wavelength excitations. 

Because the body-centered tetragonal magnetic lattice of DTN is a Bravais lattice, all expressions obtained above are applicable to DTN. The spectrum obtained using Eq.~\eqref{specs3} with the conventional parameters \eqref{olddtnpar} for DTN is presented by the solid line in Fig.~\ref{figolddtn}(a)--(c). It is seen that the agreement with experimental data is poor. At the same time the method of the spectrum calculation proposed in the present paper looks suitable for DTN. It is illustrated by Fig.~\ref{figolddtn}(d)--(f), in which the second and the third order corrections in $J/D$ to the spectrum and the first order spectrum $\e_{1\pp}$ given by Eq.~\eqref{spec0} are presented. It is seen that the third order corrections are 3--5 times smaller than the second order ones in almost the whole Brillouin zone except for the vicinity of the antiferromagnetic vector, where they are almost equal but still remain much smaller than $\e_{1\pp}$.

Thus, the theoretical description of DTN needs revision that is indicated also by recent ESR-experimental data. As it is mentioned above, the interaction between spins from different DTN sublattices is usually ignored. But the data of the recent ESR experiment \cite{12} indicate that it should be taken into consideration: one of the spectrum branches has a gap in the canted antiferromagnetic phase at $H_{c1}<H<H_{c2}$ and the optical mode is slightly split. At the same time the model \eqref{ham} without the inter-sublattice interaction has the doubly degenerate spectrum (due to two equivalent magnetic sublattices).
 
In our previous paper \cite{prev} the spin-wave approach and the magnon Bose-condensation technique (near $H_{c2}$) are used for analysis of the canted antiferromagnetic phase. It is shown that the inter-sublattice interaction of the form
\begin{eqnarray}
\label{v}
 {\cal H}_V = \sum_{i,j} V_{i,j} {\bf S}_i {\bf S}_j,
\end{eqnarray}
where index $i$ labels sites of one sublattice and $j$ labels sites of another sublattice nearest to $i$, leads to the effects observed in the ESR experiment (the gap in one of the spectrum branches and the optical mode splitting). Unfortunately the lack of experimental data near $H_{c2}$, large anisotropy and quasi-1D nature of DTN did not allow to make reliable quantitative predictions about the value of $V_{i,j}$. It is just shown in Ref.~\cite{prev} that if $V_{i,j} = V$, then $V\sim0.1$~K.

\begin{figure}
 \includegraphics[scale=0.6]{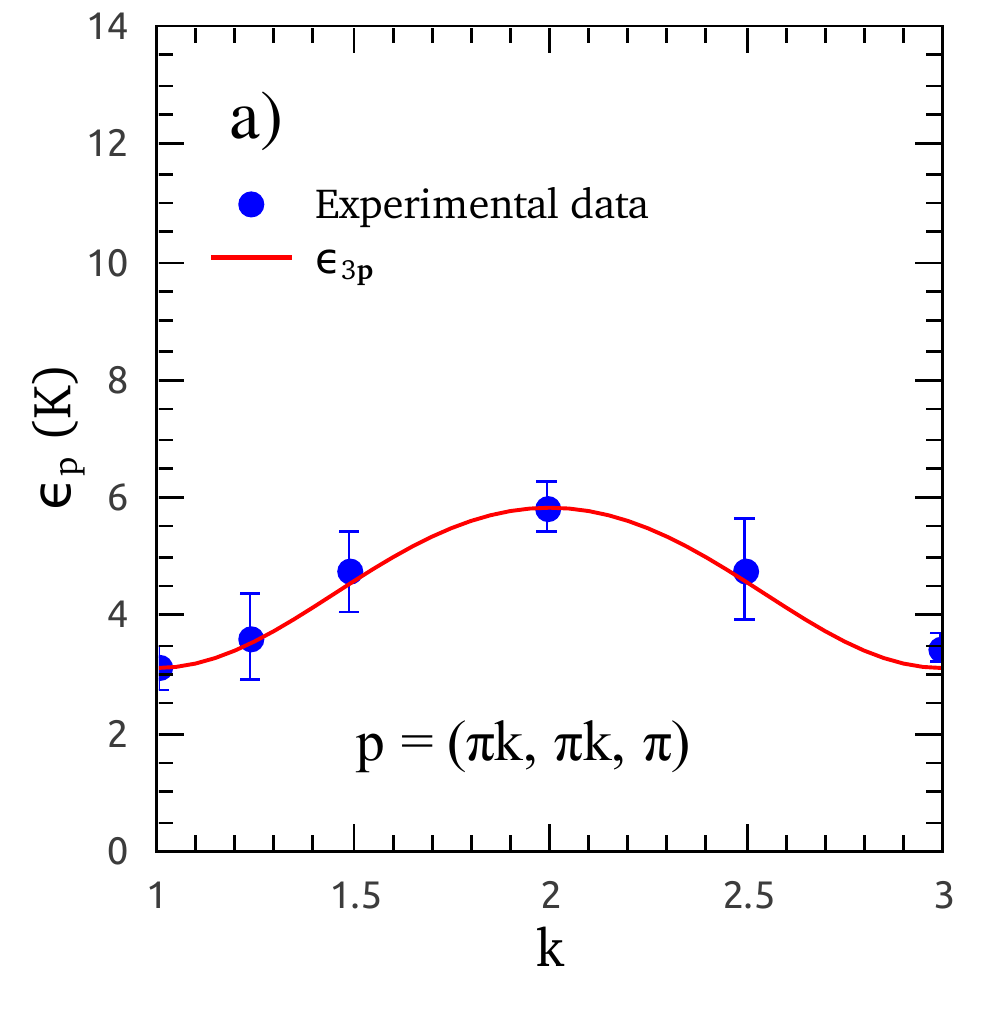}
 \includegraphics[scale=0.6]{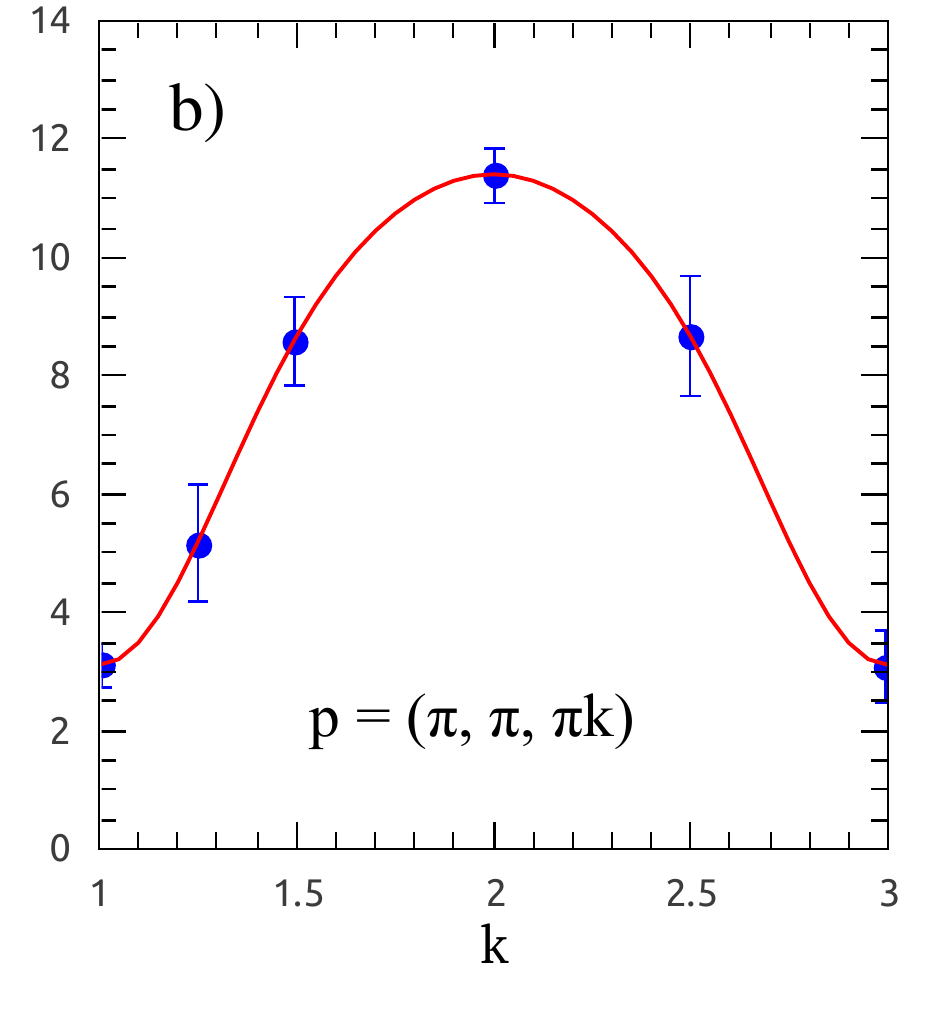}
 \includegraphics[scale=0.6]{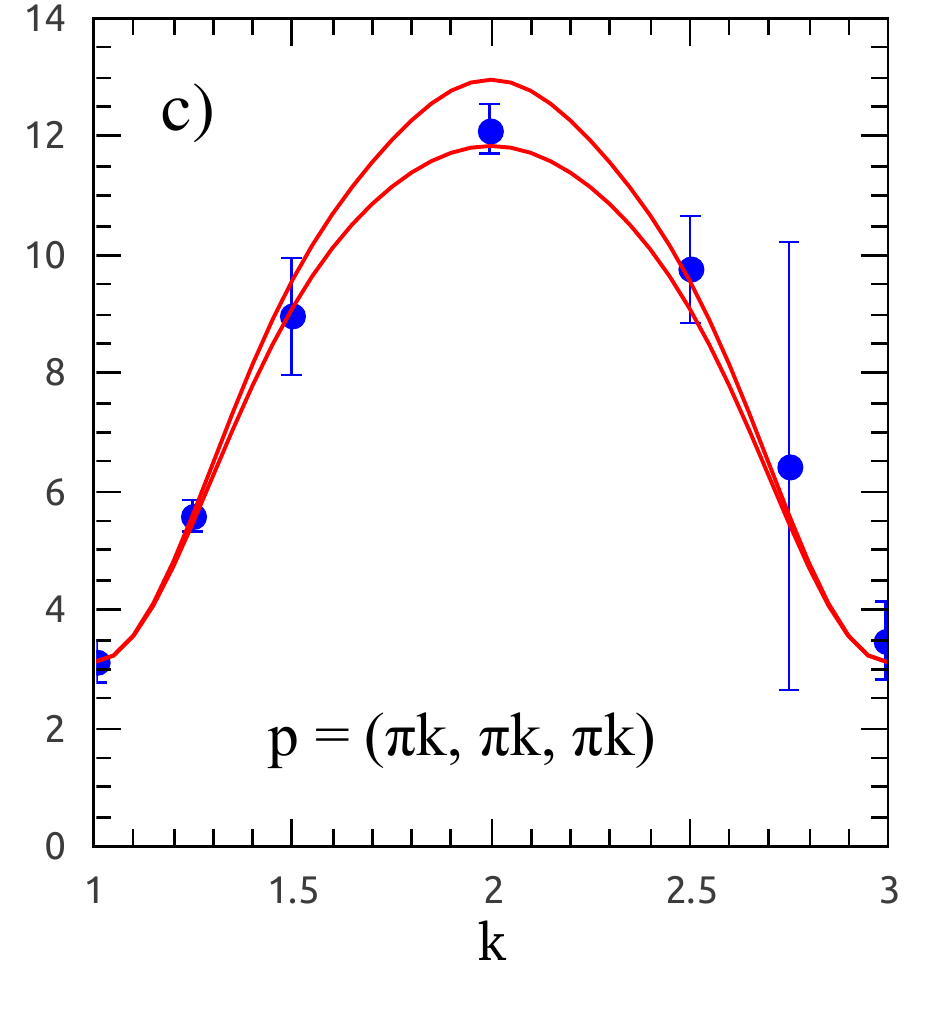}
 \caption{(Color online.) DTN spectrum along three directions in the Brillouin zone calculated using Eq.~\eqref{specs3} and parameters \eqref{ourdtnpar}. Neutron experimental data are taken from Ref.~\cite{13}.}
  \label{figourdtn}
\end{figure}

In the present paper we continue our discussion started in Ref.~\cite{prev} and fit the elementary excitation spectrum using the least square method by varying parameters $D$, $J_z$, $J_{xy}$ and $V$, while keeping the critical fields given by Eqs.~\eqref{hc1} and \eqref{hc2} to be equal to the experimentally obtained values \eqref{hcdtn}. As a result of this fit we obtain the following set of parameters that differs noticeably from the conventional one \eqref{olddtnpar}:
\begin{eqnarray}
 D &=& 7.72 \; \rm{K}, \non \\
 J_z &=& 1.86\; {\rm K}, \label{ourdtnpar} \\
 J_{xy} &=& 0.2\; {\rm K}, \non\\
 V &=& 0.1\; \rm{K}. \non
\end{eqnarray}
The spectrum obtained is presented in Fig.~\ref{figourdtn} that is in good agreement with the neutron data of Ref.~\cite{13}. The inter-sublattice interaction removes the double degeneracy of the spectrum in DTN. This splitting is zero in (a) and (b) panels of Fig.~\ref{figourdtn}, but it is clearly seen in panel (c). It should be noted that the upper branch of the spectrum in Fig.~\ref{figourdtn}(c) goes beyond the experimental error near $\pp={\bf 0}$. However, the branch splitting value $\delta\e_\pp=\e^{+}_\pp-\e^{-}_\pp$ is very small compared to $\e^{+}_\pp+\e^{-}_\pp$ and it appears to be given in DTN by a slowly convergent series in $J/D$: $\delta\e_\pp=(1.7-1.0+0.4)$~K, where the first, the second and the third terms stand for the values of corresponding corrections in $J/D$. Then, one has to take into account higher order terms in $J/D$ in order to find the small value of $\delta\e_\pp$ in DTN that is out of the scope of the present paper.

Note also that the value of $V=0.1$~K found above agrees with the estimation $V\sim0.1$~K obtained as a result of our consideration \cite{prev} of the canted antiferromagnetic phase.

\section{Conclusion}
\label{conc}

We propose the new representation \eqref{sz}, \eqref{s+} for an integer spin $S$ via bosonic operators, which is useful in describing the paramagnetic phase and transitions to magnetically ordered phases in magnetic systems with large single-ion easy-plane anisotropy. Using this representation, the diagram technique and treating the exchange interaction as a perturbation we obtain Eq.~\eqref{specs3} for the elementary excitation spectrum of the model \eqref{ham} in the paramagnetic phase in the third order of the perturbation theory (that is referred to as an expansion in terms of $J/D$ for shot). Expression \eqref{gs} is also found for the ground state energy in the third order in $J/D$. Eq.~\eqref{specs3} coincides with that obtained in Ref.~\cite{4} in the special case of a spin chain with $S=1$ and the exchange interaction between nearest neighbors only. We recover Eq.~\eqref{specs3} also at $S=1$ using simpler spin representations \eqref{sz}, \eqref{s+simple} and \eqref{sz}, \eqref{s+no}.

Comparison with numerical results obtained in Refs.~\cite{2,3} for the square lattice, $S=1$ and the antiferromagnetic exchange between nearest neighbors only shows that Eqs.~\eqref{specs3} and \eqref{gs} work better than results of other analytical methods proposed so far. In particular, Eqs.~\eqref{specs3} and \eqref{gs} work very well when $D$ is not very close to the critical value $D_{c}$, below which the antiferromagnetic phase becomes stable \cite{2,3,wong} (see Figs.~\ref{figMC}--\ref{figgs}). At $D\agt D_c$ Eq.~\eqref{specs3} poorly describes only the spectrum of long-wavelength quasiparticles. 

It is shown that Eq.~\eqref{specs3} is applicable for the spectrum analysis of the intensively studied compound $\rm NiCl_2$-$\rm 4SC(NH_2)_2$ (DTN), which is described by the model \eqref{ham} \cite{add8,chern,12,13,14,15,16,17}. We show that Eq.~\eqref{specs3} with the conventional set of parameters for DTN  \cite{14} \eqref{olddtnpar} describes the experimentally obtained spectrum \cite{13} unsatisfactorily (see Fig.~\ref{figolddtn}). The new set of parameters \eqref{ourdtnpar} is proposed for DTN which provides a good description of the experimental spectrum (see Fig.~\ref{figourdtn}) and reproduces the experimentally obtained critical fields values \cite{16} \eqref{hcdtn}. In contrast to the conventional model proposed for DTN before we take into account also the exchange interaction \eqref{v} between DTN magnetic sublattices, which becomes apparent in the recent ESR experiment \cite{12}.

In the forthcoming paper we continue analysis of the model \eqref{ham} using representations \eqref{sz}, \eqref{s+}, \eqref{s+simple} and \eqref{s+no} and consider its behavior in the vicinity of QCP $H=H_{c1}$ at $T\ne0$.

\begin{acknowledgments}

We are thankful to Prof.~A.~I.~Smirnov for stimulating discussions. This work was supported by RF President (grant MK-329.2010.2), RFBR grant 09-02-00229, and Programs "Quantum Macrophysics", "Strongly correlated electrons in semiconductors, metals, superconductors and magnetic materials" and "Neutron Research of Solids".

\end{acknowledgments}

\bibliography{LargeD} 

\end{document}

%% file: com.tex
\newcommand{\half}{\dfrac{1}{2}}

\newcommand{\NN}{\frac{1}{N}}
\newcommand{\fracNN}{\NN}

\newcommand{\beq}{\begin{eqnarray}}
\newcommand{\eeq}{\end{eqnarray}}

\newcommand{\bfig}{ \left. \begin{array}{l} }
\newcommand{\bfigor}{ \left\{ \begin{array}{ll} }
\newcommand{\efig}{ \end{array} \right. }

\newcommand{\non}{\nonumber}

\newcommand{\w}{\omega}
\newcommand{\e}{\epsilon}

\newcommand{\Sig}{\Sigma}

\newcommand{\krest}{\dagger}
\newcommand{\akrest}{a^\krest}
\newcommand{\bkrest}{b^\krest}
\newcommand{\akr}{\akrest}
\newcommand{\bkr}{\bkrest}
\newcommand{\kk}{{\bf k}}
\newcommand{\qq}{{\bf q}}
\newcommand{\pp}{{\bf p}}

\newcommand{\ppp}{{{\pp}_0}}